\documentclass[10pt,journal,twoside]{IEEEtran}
\usepackage{multirow}
\usepackage{amssymb}
\usepackage[dvips]{graphicx}
\usepackage{amsmath}
\usepackage{amsthm}
\usepackage{latexsym,bm}
\usepackage{color}
\usepackage{subfigure}
\usepackage{longtable}
\usepackage{accents}
\usepackage{cite}
\usepackage{enumerate}
\usepackage{arydshln}
\usepackage{slashbox}
\usepackage{algorithmic}
\usepackage{float}
\usepackage{subfloat}
\makeatletter
\def\widebar{\accentset{{\cc@style\underline{\mskip10mu}}}}
\def\Widebar{\accentset{{\cc@style\underline{\mskip8mu}}}}
\makeatother

\theoremstyle{plain}

\theoremstyle{definition}
\theoremstyle{definition}

\begin{document}

\title{SR-DCSK Cooperative Communication System with Code Index Modulation: A New Design for 6G New Radios}

\author{\fontsize{11pt}{\baselineskip}\selectfont {Yi Fang, {\em Member, IEEE}, Wang Chen, Pingping Chen, {\em Member, IEEE}, Yiwei Tao,\\ and Mohsen Guizani, {\em Fellow, IEEE}}
\thanks{Yi~Fang, Wang~Chen, and Yiwei~Tao are with the School of Information Engineering, Guangdong University of Technology, Guangzhou, China (e-mail: fangyi@gdut.edu.cn; wangchen\_gdut@163.com; taoyiwei0806@163.com).}
\thanks{Pingping~Chen is with School of Advanced Manufacturing, Fuzhou University, Jinjiang campus, China (e-mail: ppchen.xm@gmail.com).}
\thanks{Mohsen~Guizani is with the Department of Machine Learning, Mohamed Bin Zayed University of Artificial Intelligence (MBZUAI), Abu Dhabi, UAE. (e-mail: mguizani@ieee.org).}
}


\markboth{}%
{\MakeLowercase{\textit{}}}

\maketitle
\begin{abstract}This paper proposes a high-throughput short reference differential chaos shift keying cooperative communication system with the aid of code index modulation, referred to as CIM-SR-DCSK-CC system. In the proposed CIM-SR-DCSK-CC system, the source transmits information bits to both the relay and destination in the first time slot, while the relay not only forwards the source information bits but also sends new information bits to the destination in the second time slot.
To be specific, the relay employs an $N$-order Walsh code to carry additional ${{\log }_{2}}N$ information bits, which are superimposed onto the SR-DCSK signal carrying the decoded source information bits. Subsequently, the superimposed signal carrying both the source and relay information bits is transmitted to the destination.
Moreover, the theoretical bit error rate (BER) expressions of the proposed CIM-SR-DCSK-CC system are derived over additive white Gaussian noise (AWGN) and multipath Rayleigh fading channels. Compared with the conventional DCSK-CC system and SR-DCSK-CC system, the proposed CIM-SR-DCSK-CC system can significantly improve the throughput without deteriorating any BER performance. As a consequence, the proposed system is very promising for the applications of the 6G-enabled low-power and high-rate communication.
\end{abstract}

\section{Introduction}
\label{Introduction}

In wireless communications, multipath fading is the main factor affecting the system performance. One technique to tackle this problem is chaotic signal \cite{899921}. Especially, when the channel is time-varying or suffers from multipath propagation, the chaotic communication can exhibit excellent error performance. As the most typical and popular chaotic modulation scheme, differential chaos shift keying (DCSK) \cite{G1996Differential} has attracted much attention over the past two decades because it does not require chaos synchronization for detection. Due to the robustness against multipath fading and low-power property, DCSK has been widely applied in short-range wireless communications, e.g., wireless body area networks (WBANs) and wireless sensor networks (WSNs) \cite{4357423,8782883}.

At present, DCSK has been incorporated into more diverse communication scenarios, such as underwater acoustic communication system \cite{2019Multi}, orthogonal frequency division multiplexing (OFDM) system \cite{8424332} and ultra-wideband communication system \cite{7442517,8606201}. Moreover, some powerful error-correction codes have been combined with DCSK systems to improve performance \cite{9519519,9600574}. However, for the conventional DCSK system \cite{G1996Differential}, each data frame is divided into two time slots, one time slot is used to transmit reference-chaotic signals, and the other time slot is used to transmit information-bearing signals. This structure results in low data rate, low spectral efficiency and waste of energy. Thereby, several feasible approaches have been proposed to address the above drawbacks. In particular, a short reference DCSK (SR-DCSK) scheme has been presented in \cite{7370796} to effectively improve the data rate and energy efficiency by reducing the frame duration. An enhanced DCSK scheme has been designed in \cite{780045} to increase date rates and save transmission energy. Additionally, a high-efficiency DCSK scheme (HE-DCSK) has been presented in \cite{6184294} to boost the energy efficiency by recycling reference sample. In \cite{6560492}, a multi-carrier DCSK (MC-DCSK) scheme, which can be seen as a parallel extension of the conventional DCSK scheme, has been conceived to achieve high data rate and energy efficiency. Moreover, the constellation-aided $M$-ary DCSK system has been appeared to another effective method to achieve high data rate and high spectral efficiency \cite{7109922,7572953,7579619}.

Index modulation (IM) is a new emerging technique that can improve data rate and energy efficiency of wireless communication systems \cite{8425986,8734769}. In IM technique, various transmission resources such as spreading code, time slot, sub-carrier and transmit antenna can be used to carry information \cite{7929332,9380189}. The code index modulation (CIM) technique was first applied to direct-spread spectrum-sequence communication system in \cite{6994807,7317808} to achieve the goal of increasing data rate. This technique also has potential to overcome the drawbacks of low data rate and energy efficiency of DCSK system. In recent years, there have been some research works touching upon the joint design of IM and DCSK \cite{9361588,9260203,9761226}. Especially, a new CIM-aided SR-DCSK that uses a Walsh code to carry additional information bits, called CIM-DCSK, has been developed in \cite{8290668} to obtain high data rate. Moreover,
a hybrid modulation scheme integrating pulse position modulation (PPM) with DCSK has been presented in \cite{8468068}, which exploits the activation pattern of time slots to carry additional information bits. In \cite{7590017}, a carrier index modulation DCSK (CI-DCSK) scheme has been devised, which carries additional information bits by activating different sub-carriers. Furthermore, a dual-mode DCSK scheme with index modulation (DM-DCSK-IM) has been proposed in \cite{8721243} to exploit both inactive and active time slots to convey information bits, hence significantly boosting of data rate of the PPM-DCSK scheme.

As another promising anti-multipath-fading technique, spatial diversity has also attract much attention \cite{proakis2001digital,9142447}. Actually, the basic principle of spatial diversity is to mitigate the effect of multipath fading by transmitting redundant signal information via multiple antennas.
Inspired by the advantage of spatial diversity, the joint design of DCSK and multiple-input multiple-output (MIMO) technique has been investigated in \cite{5937879,6298988}. Considering the cost and size of the equipments, the deployment of multiple antennas is desired in actual implementations. Hence, the cooperative communication, which realizes transmit diversity via the deployment of a relay between the source and destination, has been proposed as an alternative approach to combat the multipath fading \cite{liu2009cooperative}.
In the conventional cooperative communication system, the relay generally adopts amplify-and-forward (AF) or decode-and-forward (DF) protocol to transmit source information to the destination \cite{9730788,7314985}. In particular, in the half-duplex mode, the relay preferentially helps the source to forward the information and delays its own information transmission to the destination.
As such, the relay needs to transmit its own information to the destination at the cost of extra time slots. To improve the transmission throughput, IM technique can be applied to cooperative communication. For instance, a distributed OFDM cooperative system with the aid of IM has been devised in \cite{9205985}, in which the information bits of source and relay can be transmitted simultaneously to the destination in each transmission period.

On the other hand, it has been demonstrated in \cite{5629387,8036271,9213137,9142258} that combining DCSK with cooperative communication can greatly improve the reliability of wireless communication systems. 
In particular, a DCSK-based cooperative communication (DCSK-CC) system has been presented in \cite{5537139}, but it has a relatively lower data rate compared to the conventional DCSK system because an extra time slot is allocated to the relay for transmission. In \cite{7577838}, an efficient transmission scheme for DCSK-CC system has been devised, which exploits an efficient partial-sequence cooperative communication  (PS-CC) scheme to solve the problem of low data rate.

To preserve the advantages of the DCSK-CC system while improving the data rate, we propose a high-throughput CIM-aided SR-DCSK cooperative communication system, referred to as  CIM-SR-DCSK-CC system, in this paper. The contributions in this paper are summarized as follows:
\begin{enumerate}
\item[1)]
In the proposed CIM-SR-DCSK-CC system, the source transmits information bits to the relay and the destination in the first time slot, while the relay transmits its own information bits and source information bits simultaneously to the destination in the second time slot. More specifically, we exploit the SR-DCSK to realize the communication between the source and destination, and employ the CIM-SR-DCSK to guarantee the communication request between the relay and destination without wasting additional time slots and energy.

\item[2)]
We derive the theoretical bit error rate (BER) formulas of the proposed CIM-SR-DCSK-CC system over additive white Gaussian noise (AWGN) and multipath Rayleigh fading channels and verify their accuracy via Monte-Carlo simulations.

\item[3)]
We compare the performance of the proposed CIM-SR-DCSK-CC system with other DCSK cooperative communication systems, i.e., DCSK-CC system and SR-DCSK-CC system. The results demonstrate that the proposed system can achieve similar BER performance as the two counterparts over AWGN and multipath Rayleigh fading channels, but achieves a significant improvement in terms of throughput.
\end{enumerate}

In consequence, the proposed system has great potential to realize low-power high-rate transmissions in 6G-enabled applications.

The remainder of this paper is structured as follows. Section~\ref{sect:system model} briefly reviews conventional SR-DCSK system and presents the proposed CIM-SR-DCSK-CC system. Section~\ref{sect:Performance analysis of the system} analyzes the BER and throughput performance of the CIM-SR-DCSK-CC system. Section~\ref{sect:simulation results and discussion} presents simulation results and discussions. Finally, Section~\ref{sect:Conclusions} draws the conclusion.

\section{System Model}
\label{sect:system model}
This section first introduces the basic principle of SR-DCSK system, and then describes the proposed high-throughput CIM-SR-DCSK-CC system model.

\subsection{SR-DCSK System} \label{sect:SR-DCSK system}
\begin{figure}[h]
\center
\subfigure[]{ \label{fig:subfig:1a}
\includegraphics[width=2.82in,height=1.12in]{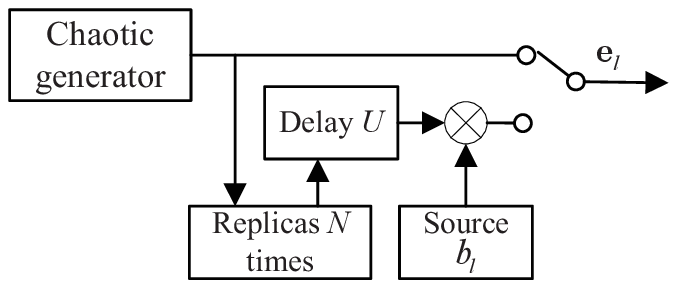}}
\vspace{-2mm}
\subfigure[]{ \label{fig:subfig:1b}
\includegraphics[width=2.82in,height=1.12in]{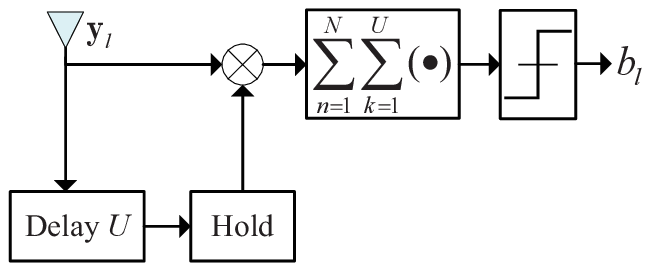}}
\vspace{-2mm}
\subfigure[]{ \label{fig:subfig:1c}
\includegraphics[width=3.2in,height=0.6in]{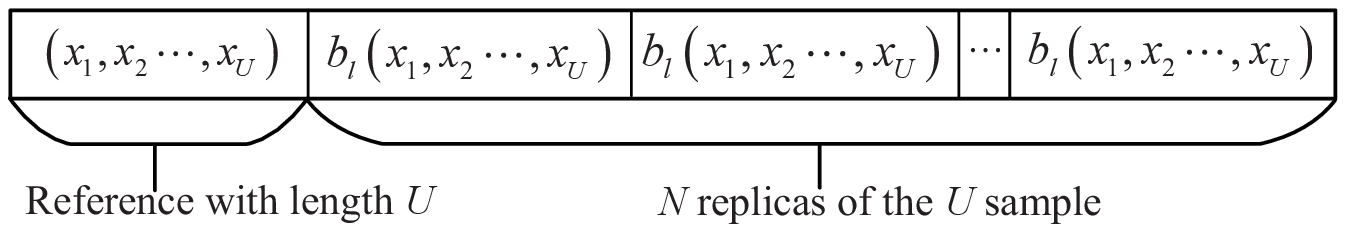}}
\caption{Structures of (a) SR-DCSK transmitter; (b) SR-DCSK receiver; and (c) SR-DCSK signal.}
\label{fig:fig1}
\end{figure}
The block diagram of the SR-DCSK transceiver is shown in Figure~\ref{fig:fig1}\subref{fig:subfig:1a} and Figure~\ref{fig:fig1}\subref{fig:subfig:1b}~\cite{7370796}. As seen, the transmitted signal consists of a reference-chaotic signal and an information-bearing signal. The second-order Chebyshev polynomial function (CPF), i.e., ${{x}_{k+1}}=1-2x_{k}^{2}$ is used to generate a $U$-length reference-chaotic signal $\mathbf{x}$, where $\mathbf{x}=\left[ {{x}_{1}},{{x}_{2}}\cdots ,{{x}_{U}} \right]$. The $\beta$-length information-bearing signal is generated by copying the reference-chaotic signal $N$ times, where $\beta =N\cdot U$. Thereby, the structure of an SR-DCSK signal can be obtained as Figure~\ref{fig:fig1}\subref{fig:subfig:1c}, and spreading factor is defined as $SF=U+\beta =(N+1)U$. For each transmitted symbol
${{b}_{l}}\in \left\{+1,-1 \right\}$, the transmitted signal of the $l$-$\rm{th}$ symbol can be expressed as
\begin{align} \label{eq:1func}
 {{\mathbf{e}}_{l}}=[\underbrace{\mathbf{x}}_{\rm reference},\ \underbrace{{{b}_{l}}{{\mathbf{I}}_{N}}\otimes \mathbf{x}}_{\rm information\text{-}bearing}],
\end{align}
where ${{\mathbf{I}}_{N}}={{\left[ 1,\cdots ,1 \right]}_{1\times N}}$ is a unit vector of length $N$ and $\otimes$ represents the Kronecker product.

The transmitted signal passes through a wireless channel and then yields the received signal ${{\mathbf{y}}_{l}}$. 
The received  reference-chaotic signal is correlated with $N$ replicas of information-bearing signal. Then, we can obtain $N$ independent correlation values and get the decision metric by summing these correlation values. Finally, the transmitted symbol ${{b}_{l}}$ can be estimated by comparing the decision metric with a zero threshold.
\begin{table*}[htbp]
\centering
\caption{Illustration of the mapping rule between index bits and index symbol ${{a}_{l}}$ (i.e., row vector of Walsh code) in the proposed CIM-SR-DCSK-CC system.}
\label{tab:1}
\setlength{\tabcolsep}{5mm}\begin{tabular}{ccc}
\hline
\textbf{ Index Bits \qquad} & \textbf{Index Symbol ${{a}_{l}}$ } & \textbf{Selected Walsh-Code Vector} \\\hline
{${{\left[ \begin{matrix}
    0 &\cdots & 0  \\
\end{matrix} \right]}_{1\times {{m}_{c}}}}$
} & 1 & {${{\mathbf{w}}_{1}}={{\left[ \begin{matrix}
   +1 & +1 & \cdots  & +1 & +1  \\
\end{matrix} \right]}_{1\times N}}$

}\\
 {${{\left[ \begin{matrix}
    0  & \cdots  & 1  \\
\end{matrix} \right]}_{1\times {{m}_{c}}}}$
} & 2 & {${{\mathbf{w}}_{2}}={{\left[ \begin{matrix}
   +1 & -1 & \cdots  & +1 & -1  \\
\end{matrix} \right]}_{1\times N}}$

}\\
{$\vdots $ \ \ \qquad} & $\vdots $ & {\quad $\vdots\  $
}\\
 {${{\left[ \begin{matrix}
   1 & \cdots  & 1  \\
\end{matrix} \right]}_{1\times {{m}_{c}}}}$
} & $N$ & {${{\mathbf{w}}_{N}}={{\left[ \begin{matrix}
   +1 & -1 & \cdots  & -1 & +1  \\
\end{matrix} \right]}_{1\times N}}$
}\\ \hline
\end{tabular}
\end{table*}

\subsection{Proposed CIM-SR-DCSK-CC System}\label{sect: a CIM-SR-DCSK-CC system model}
\vspace{5mm}
We consider a half-duplex cooperative communication system including a source node S, a destination node D, and a relay node R. In particular, we assume that both S and R need to transmit information bits to D, and R adopts the DF protocol. In this system, the information bit streams are composed of both modulated bit and index bits. The modulated bit and index bits are transmitted by S and R, respectively. There are two time slots in each transmission period. Specifically, the transmission period of the CIM-SR-DCSK-CC system is ${\left( U+\beta \right){{N}_{p}}{{T}_{c}}\times 2}$, where ${{N}_{p}}$ is the number of information bit streams sent in each transmission period and ${{T}_{c}}$ is the chip time.
In the first time slot, we take the transmission of the $l$-$\rm{th}$ information bit stream as an example. In this case, the modulated bit is first mapped to modulated symbol ${{b}_{l}}$ at S. Then, ${{b}_{l}}$ is carried by the SR-DCSK signal and transmitted to R and D simultaneously. In the second time slot, R adopts the DF protocol to recover the modulated symbol $b_{l}$ and transmit a new chaotic signal to D. The new chaotic signal is formed by combining the SR-DCSK signal carrying modulated symbol $b_{l}$ and a specific row vector of Walsh code. In particular, the choice of specific row vector is determined by index symbol ${{a}_{l}} \in \{1,2,\ldots, N\}$, which is the mapped by the index bits. Moreover, the number ${m}_{c}$ of index bits is related to the number $N$ of replicas of the SR-DCSK signal, i.e., ${{m}_{c}}={{\log }_{2}}N$.
\vspace{1mm}

\begin{figure}[htbp]
\center
\includegraphics[width=3.25in,height=1.50in]{{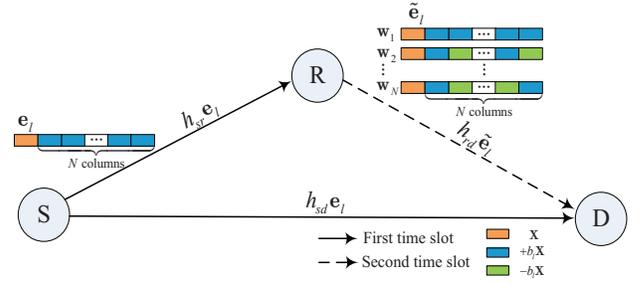}}
\vspace{5mm}
\caption{The proposed CIM-SR-DCSK-CC system model.}
\label{fig:fig2}
\vspace{5mm}
\end{figure}

As a further advance, Table \ref{tab:1} illustrates the mapping rule between the index bits and index symbol ${{a}_{l}}$ (i.e., row vector of Walsh code) in the proposed CIM-SR-DCSK-CC system.
Finally, D processes the signals transmitted from S and R to demodulate the overall information bits. Figure~\ref{fig:fig2} depicts the proposed CIM-SR-DCSK-CC system model.

The transmitter of S is shown in Figure~\ref{fig:fig1}\subref{fig:subfig:1a} and the transmitted signal ${{e}_{sr(sd)}}$ is expressed as Eq.~(\ref{eq:1func}). In the first time slot, after passing through the multipath Rayleigh fading channels, the signals received by R and D can be respectively given by
\begin{align} \label{eq:2func}
{{y}_{sr,k}}&=\sqrt{\frac{{{P}_{S}}}{d_{sr}^{\alpha }}}\sum\limits_{l=1}^{{{L}_{sr}}}{{{h}_{sr,l}}}{{e}_{sr,k-{{\tau }_{sr,l}}}}+{{n}_{sr,k}},
 \end{align}
 \begin{align}
{{y}_{sd,k}}&=\sqrt{\frac{{{P}_{S}}}{d_{sd}^{\alpha }}}\sum\limits_{l=1}^{{{L}_{sd}}}{{{h}_{sd,l}}}{{e}_{sd,k-{{\tau }_{sd,l}}}}+{{n}_{sd,k}},
 \end{align}
where ${{P}_{S}}$ is the transmit power at S, ${{d}_{sr(sd)}}$ is the distance between S and R (D), ${{L}_{sr(sd)}}$ is the number of channel paths, $\alpha$ is the path loss coefficient, ${{h}_{sr(sd),l}}$ and ${{\tau }_{sr(sd),l}}$ represent the channel coefficient and delay of the $l$-$\rm{th}$ path for S to R (D), respectively. Besides, the ${{n}_{sr(sd),k}}$ denotes the AWGN with zero mean and variance ${{N}_{0,sr(sd)}}/2$ for S$\rightarrow$R (D) link.

On the one hand, the receiver structure at R is shown in Figure~\ref{fig:fig1}\subref{fig:subfig:1b}. Further, the decision metric of S$\rightarrow$R link can be expressed as
 \vspace{-2mm}
\begin{align}\label{eq:4func}
   {{Z}_{sr}}=&\sum\limits_{n=1}^{N}{\sum\limits_{k=1}^{U}{{{y}_{sr,k}}\times{{y}_{sr,k+nU}}}} \nonumber \\
  =&\sum\limits_{n=1}^{N}{}\sum\limits_{k=1}^{U}{}\left( {{b}_{l}}\frac{{{P}_{S}}}{d_{sr}^{\alpha }}\sum\limits_{l=1}^{{{L}_{sr}}}{}h_{sr,l}^{2}x_{k-{{\tau }_{sr}}}^{2} \right. \nonumber \\
 & +\sqrt{\frac{{{P}_{S}}}{d_{sr}^{\alpha }}}\sum\limits_{l=1}^{{{L}_{sr}}}{}{{h}_{sr,l}}{{x}_{k-{{\tau }_{sr}}}}{{n}_{sr,k+nU}}  \nonumber\\
 & \left.\! +\!{{b}_{l}}\sqrt{\frac{{{P}_{S}}}{d_{sr}^{\alpha }}}\!\sum\limits_{l=1}^{{{L}_{sr}}}{}\!{{h}_{sr,l}}{{x}_{k\!-{{\tau }_{sr}}}}{{n}_{sr,k}}\!+\!{{n}_{sr,k}}{{n}_{sr,k\!+nU}} \right)\!.
\end{align}
Hence, if ${{Z}_{sr}}>0$, the symbol $b_{l}$ is estimated as $b_{l}=+1$; otherwise, $b_{l}$ is estimated as $b_{l}=-1$.

\begin{figure*}[htbp]
\center
\includegraphics[scale=0.836]{{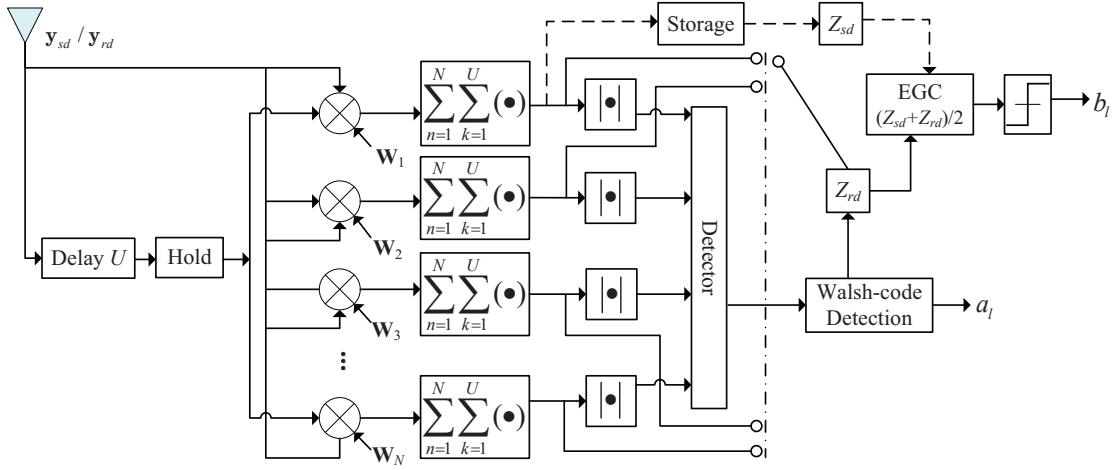}}
\caption{Structure of receiver at D in the proposed CIM-SR-DCSK-CC system.}
\label{fig:fig3}
\end{figure*}

On the other hand, the structure of receiver at D is shown in Figure~\ref {fig:fig3}. In the first time slot, we only need to perform the operation in the first branch of the detector to process the received signal of S$\rightarrow$D link, then get the decision metric and store it. This operation is equivalent to SR-DCSK modulation, because the first row of Walsh code is all-ones vector, i.e., ${{\mathbf{w}}_{1}}={{\left[ +1,\cdots ,+1 \right]}_{1\times N}}$. Hence, the decision metric of S$\rightarrow$D link can be expressed as
\begin{align}\label{eq:5func}
   {{Z}_{sd}}=&\sum\limits_{n=1}^{N}{\sum\limits_{k=1}^{U}{{{y}_{sd,k}}\times{{y}_{sd,k+nU}}}} \nonumber \\
  =&\sum\limits_{n=1}^{N}{}\sum\limits_{k=1}^{U}{}\left( {{b}_{l}}\frac{{{P}_{S}}}{d_{sd}^{\alpha }}\sum\limits_{l=1}^{{{L}_{sd}}}{}h_{sd,l}^{2}x_{k-{{\tau }_{sd}}}^{2} \right. \nonumber \\
 & +\sqrt{\frac{{{P}_{S}}}{d_{sd}^{\alpha }}}\sum\limits_{l=1}^{{{L}_{sd}}}{}{{h}_{sd,l}}{{x}_{k-{{\tau }_{sd}}}}{{n}_{sd,k+nU}}  \nonumber\\
 & \left.\! +\!{{b}_{l}}\sqrt{\frac{{{P}_{S}}}{d_{sd}^{\alpha }}}\!\sum\limits_{l=1}^{{{L}_{sd}}}{}\!{{h}_{sd,l}}{{x}_{k\!-{{\tau }_{sd}}}}{{n}_{sd,k}}\!+\!{{n}_{sd,k}}{{n}_{sd,k\!+nU}}\! \right)\!.
\end{align}

\begin{figure}[h]
\centering
\subfigure[\hspace{-0.0cm}]{ \label{fig:subfig:4a}
\includegraphics[scale=0.72]{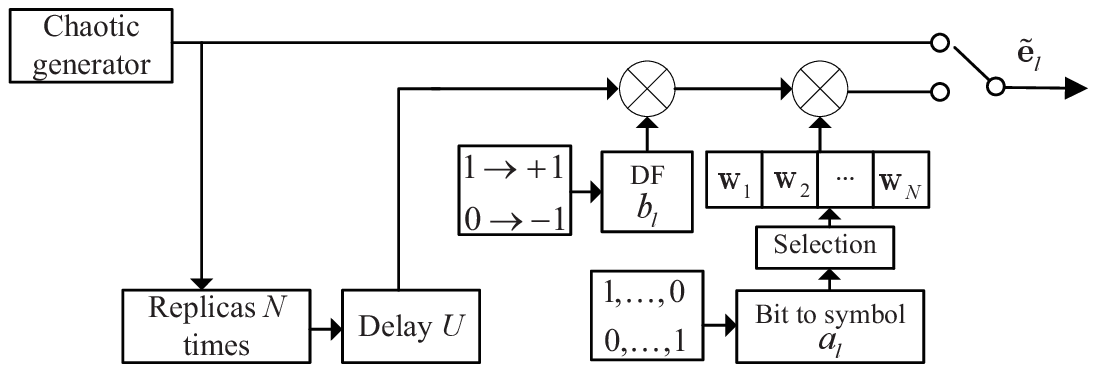}}\\
\subfigure[\hspace{-0.0cm}]{ \label{fig:subfig:4b}
\includegraphics[width=3.25in,height=0.6in]{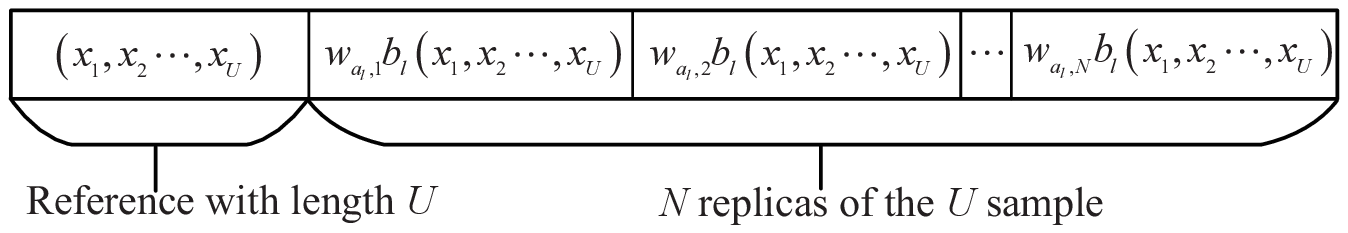}}
\caption{Structures of (a) transmitter and (b) transmitted signal at R in the proposed CIM-SR-DCSK-CC system.}
\label{fig:fig4}
\end{figure}

In the second time slot, R transmits the decoded source information bit and its own information bits to D via the combination of SR-DCSK signal and a specific row vector of Walsh code.
The structures of the transmitter and transmitted signal at R are shown in Figure~\ref{fig:fig4}\subref{fig:subfig:4a} and Figure~\ref{fig:fig4}\subref{fig:subfig:4b}, respectively. To be specific, the transmitted signal of the $l$-$\rm{th}$ symbol is written as
 \begin{align} \label{eq:6func}
 {{\mathbf{\tilde{e}}}_{l}}=[\underbrace{\mathbf{x}}_{\rm{reference}}, \ \underbrace{{{b}_{l}}{{\mathbf{w}}_{{{a}_{l}},N}}\otimes \mathbf{x}}_{\rm{information-bearing}}],
\end{align}
where ${{\mathbf{w}}_{{{a}_{l}},N}}$ represents the ${a}_{l}$-$\rm{th}$ row vector of $N$-order Walsh code.  After passing through the multipath Rayleigh fading channels, the signal received by D is yielded as
 \begin{align} \label{eq:7func}
{{y}_{rd,k}}=\sqrt{\frac{{{P}_{{R}}}}{d_{rd}^{\alpha }}}\sum\limits_{l=1}^{{{L}_{rd}}}{{{h}_{rd,l}}}{{\tilde{e}}_{rd,k-{{\tau }_{rd,l}}}}+{{n}_{rd,k}},
 \end{align}
where  ${{P}_{R}}$ is the transmit power at R, ${{d}_{rd}}$ is the distance between R and D, ${{L}_{rd}}$ is the number of channel paths, ${{h}_{rd,l}}$ and ${{\tau }_{rd,l}}$ represent the channel coefficient and delay of the $l$-$\rm{th}$ path for R to D, respectively. Besides, ${{n}_{rd,k}}$ denotes the AWGN with zero mean and variance ${{N}_{0,rd}}/2$ for R$\rightarrow$D link.

Then, we need to perform the operations of all $N$ detector branches to process the received signal of R$\rightarrow$D link, as shown in Figure~\ref {fig:fig3}.  Assume that $\hat{m} \triangleq  {{a}_{l}}$ is the estimated index symbol output by the detector. If $m\ne \hat{m}$, the output of the ${m}$-$\rm{th}$~$(m=1,2,\ldots,N)$ branch is formulated as
\begin{align} \label{eq:8func}
   {{Z}_{m}}=&\sum\limits_{n=1}^{N}{{{w}_{m,n}}\times\sum\limits_{k=1}^{U}{{{y}_{rd,k}}\times{{y}_{rd,k+nU}}}} \nonumber\\
  =&\sum\limits_{n=1}^{N}{\sum\limits_{k=1}^{U}{{}}\Bigg(\sqrt{\frac{{{P}_{R}}}{d_{rd}^{\alpha }}}\sum\limits_{l=1}^{{{L}_{rd}}}\!{{{h}_{rd,l}}}{{x}_{k\!-{{\tau }_{rd}}}}{{n}_{rd,k\!+nU}}{{w}_{m,n}} \Bigg.} \nonumber\\
 & \Bigg. +{{n}_{rd,k}}{{n}_{rd,k+nU}}{{w}_{m,n}} \Bigg).
\end{align}
Otherwise, if $m=\hat{m}$, the corresponding output becomes
\begin{align}\label{eq:9func}
   {{Z}_{{\hat{m}}}}=&\sum\limits_{n=1}^{N}{{{w}_{\hat{m},n}}\times\sum\limits_{k=1}^{U}{{{y}_{rd,k}}\times{{y}_{rd,k+nU}}}}\nonumber \\
  =&\sum\limits_{n=1}^{N}{{}}\sum\limits_{k=1}^{U}{{}}\left( {{b}_{l}}\sqrt{\frac{{{P}_{R}}}{d_{rd}^{\alpha }}}\sum\limits_{l=1}^{{{L}_{rd}}}{{{h}_{rd,l}}}{{x}_{k-{{\tau }_{rd}}}}{{n}_{rd,k}} \right. \nonumber\\
 & +\!{{b}_{l}}\frac{{{P}_{R}}}{d_{rd}^{\alpha }}\sum\limits_{l=1}^{{{L}_{rd}}}{h_{rd,l}^{2}x_{k-{{\tau }_{rd}}}^{2}}\!+\!{{n}_{rd,k}}{{n}_{rd,k+nU}}{{w}_{\hat{m},n}} \nonumber\\
  &\left. +\sqrt{\frac{{{P}_{R}}}{d_{rd}^{\alpha }}}\sum\limits_{l=1}^{{{L}_{rd}}}{{{h}_{rd,l}}}{{x}_{k-{{\tau }_{rd}}}}{{n}_{rd,k+nU}}{{w}_{\hat{m},n}} \right).
\end{align}


To recover the index bits, the index symbol ${{a}_{l}}$ can be estimated by comparing the absolute values of Eq.~(\ref{eq:8func}) and Eq.~(\ref{eq:9func}), i.e.,
\begin{align} \label{eq:10func}
&{{a}_{l}}=\arg \underset{\!m=1,\ldots ,N}{\mathop{\!\max }}\,\left( \left| {{Z}_{m}} \right| \right).
\end{align}
Hence, the decision metric of R$\rightarrow$D link can be written as ${{Z}_{rd}}={{Z}_{{{a}_{l}}}}$. Furthermore, to recover the modulated bit, an equal-gain combiner (EGC)\footnote{The proposed CIM-SR-DCSK-CC system can achieve desirable performance without requiring any channel state information (CSI), thus it is appropriate to employ a simpler combiner, e.g., equal-gain combiner (EGC). If the maximum ratio combiner (MRC) is employed, the system requires to estimate the CSI at receiver, which significantly increases the implementation complexity.}\ is utilized to process the decision metrics of S$\rightarrow$D and R$\rightarrow$D links, the modulated symbol ${{b}_{l}}$ can be estimated as
\begin{align}  \label{eq:11func}
  {{b}_{l}}=\left\{ \begin{matrix}
   +1  \\
   -1  \\
\end{matrix} \right.\begin{matrix}
   \ ({{Z}_{sd}}+{{Z}_{rd}})/2> 0  \\
   \rm otherwise
\end{matrix}.
\end{align}

\section{Performance Analysis of Proposed CIM-SR-DCSK-CC System}\label{sect:Performance analysis of the system}
\subsection{BER Performance}

\subsubsection{Formulation of System BER}\label{sect:BER of the system}

In the CIM-SR-DCSK-CC system, the overall BER is determined by the BER of index bits and the BER of modulated bit. Hence, the BER expression of the system can be written as
\begin{align} \label{eq:12func}
{{P}_{\rm{sys}}}=\frac{{{m}_{c}}}{ {{m}_{c}}+1 }{{P}_{\rm{cim}}}+\frac{1}{ {{m}_{c}}+1}{{{P}_{\rm{mod}}}},
\end{align}
where ${{P}_{\rm{cim}}}$ and ${{P}_{\rm{mod}}}$ are defined as the BERs of the index bits and modulated bit, respectively.
\subsubsection{Derivation of ${P}_{\rm{{cim}}}$}\label{sect:Derivation of ${P}_{{cim}}$}
${{P}_{\rm{cim}}}$  is related to the error probability of Walsh-code detection. More specifically, the probability of detecting any of the $N-1$ incorrect row vectors of a Walsh code is the same. Then, the expectation of the number of error bits is expressed as
\begin{align} \label{eq:13func}
Q=\sum\limits_{i=1}^{{{m}_{c}}}{i}\frac{C_{{{m}_{c}}}^{i}}{N-1},
\end{align}
where $C_{a}^{b}=\frac{a!}{b!(a-b)!}$. Hence, the BER of index bits is calculated as
\begin{align} \label{eq:14func}
{{P}_{\rm{cim}}}=\frac{Q}{{{m}_{c}}}{{P}_{\rm{ed}}},
\end{align}
where ${{P}_{\rm{ed}}}$ is defined as the error probability of Walsh-code detection. To obtain ${P}_{\rm{{ed}}}$, we assume that the modulated symbol $b_{l}={+1}$ and the index symbol ${{a}_{l}}=\hat{m}$. Thus, the mean and the variance of ${{Z}_{{\hat{m}}}}$ and ${{Z}_{m}}$ in the R$\rightarrow$D link are respectively formulated as
\begin{subequations}\label{eq:15func}
\begin{align}
{{\mu }_{1}}&=E\left\{ {{Z}_{{\hat{m}}}} \right\}={{b}_{l}}\sum\limits_{l=1}^{{{L}_{rd}}}{h_{rd,l}^{2}}\frac{N{{P}_{R}}{{E}_{s}}}{\left( 1+N \right)d_{rd}^{\alpha }},
\end{align}
\begin{align}
{{\mu }_{2}}&=E\left\{ {{Z}_{m}} \right\}=0,
\end{align}
\begin{align}
\sigma _{1}^{2}=\!V\!\left\{ {{Z}_{{\hat{m}}}} \right\}\!=\!\sum\limits_{l=1}^{{{L}_{rd}}}\!{h_{rd,l}^{2}}\frac{{{P}_{R}}{{E}_{s}}{{N}_{0,rd}}}{d_{rd}^{\alpha }}\!\left(\! \frac{N}{2}\!+\!\frac{NU}{4{{\gamma }_{rd}}} \right)\!,
\end{align}
\begin{align}
\sigma _{2}^{2}=V  \left\{{{Z}_{m}} \right\}=&\sum\limits_{l=1}^{{{L}_{rd}}}{h_{rd,l}^{2}}\frac{{{P}_{R}}{{E}_{s}}{{N}_{0,rd}}}{d_{rd}^{\alpha }} \nonumber\\
&\times\underbrace{ \left( \frac{N}{2(1\!+N)}+\frac{NU}{4{{\gamma }_{rd}}} \right)}_{\lambda },
\end{align}
\end{subequations}
where ${{E}_{s}}$ is defined as the average symbol energy of the system, i.e., ${{E}_{s}}\!=\!\left( 1+N \right)UE\left\{ x_{k}^{2} \right\}$, and ${{\gamma }_{rd}}=\sum\limits_{l=1}^{{{L}_{rd}}}{\frac{h_{rd,l}^{2}{{P}_{R}}{{E}_{s}}}{d_{rd}^{\alpha }{{N}_{0,rd}}}}$ is the instantaneous signal-to-noise ratio (SNR) of R$\rightarrow$D link.

The $\left| {{Z}_{{\hat{m}}}} \right|$ and $\left| {{Z}_{m}} \right|$  are random variables following identical folded normal distribution \cite{8290668,Michail2014On}. Hence, the probability density function (PDF) of $\left| {{Z}_{{\hat{m}}}} \right|$ and cumulative distribution function of $\left| {{Z}_{m}} \right|$ can be written as
\begin{subequations}
\begin{align} \label{eq:16func}
   &{{f}_{{ }|\!\!\text{ }{{Z}_{{\hat{m}}}}\text{ }\!\!|\!\!\text{ }}}(x)\!=\!\!\frac{1}{\sqrt{2\pi \sigma _{\left| {{Z}_{{\hat{m}}}} \right|}^{2}}}\left\{ \!{{e}^{-\frac{{{(\!x+\!\mu _{\left| {{Z}_{{\hat{m}}}} \right|}^{{}})}^{2}}}{2\sigma _{\left| {{Z}_{{\hat{m}}}} \right|}^{2}}}}\!+\!{{e}^{-\frac{{{(\!x-\!\mu _{\left| {{Z}_{{\hat{m}}}} \right|}^{{}})}^{2}}}{2\sigma _{\text{ }\!\!\!\text{ }\left| {{Z}_{{\hat{m}}}} \right|}^{2}}}} \right\}\!,\\
  &{{F}_{|{{Z}_{m}}|}}(y)\!=\!\operatorname{erf}(\frac{y}{\sqrt{2\sigma _{2}^{2}}}),
\end{align}
\end{subequations}
where the $ {\rm erf}(x)$ is error function. Subsequently, the mean and variance of $\left| {{Z}_{\hat{m}}} \right|$ are calculated by
\begin{subequations}
 \begin{align} \label{eq:17func}
  {{\mu }_{\left| {{Z}_{{\hat{m}}}} \right|}}=&\sqrt{\frac{2}{\pi }}{{\sigma }_{1}}{{e}^{-\frac{\mu _{1}^{2}}{2\sigma _{1}^{2}}}}-{{\mu }_{1}}\text{erf}(-\sqrt{\frac{\mu _{1}^{2}}{2{{\sigma }_{1}}}}) \nonumber \\
  =&\sqrt{\sum\limits_{l=1}^{{{L}_{rd}}}{}h_{rd,l}^{2}{{E}_{s}}{{N}_{0,rd}}}\Psi,
\end{align}
\begin{align}
 \sigma _{\left| {{Z}_{{\hat{m}}}} \right|}^{2}=&\mu _{1}^{2}+\sigma _{1}^{2}-\mu _{\left| {{Z}_{{\hat{m}}}} \right|}^{2}\nonumber \\
  =&\sum\limits_{l=1}^{{{L}_{rd}}}{}h_{rd,l}^{2}{{E}_{s}}{{N}_{0,rd}}\nonumber \\
&\times\! \underbrace{\left( \frac{{{P}_{R}}}{d_{rd}^{\alpha }}\left( \frac{{{N}^{2}}{{\gamma }_{rd}}}{{{(1+N)}^{2}}}+\frac{N}{2}+\frac{NU}{4{{\gamma }_{rd}}} \right)-{{\Psi }^{2}} \right)}_{\eta },
\end{align}
\end{subequations}
where
\begin{align}
  \Psi = &\sqrt{\frac{\frac{{{P}_{R}}}{d_{rd}^{\alpha }}N+\frac{NU}{2{{\gamma }_{rd}}}}{\pi }}{{e}^{-\frac{2N\gamma _{rd}^{2}}{{{(1+N)}^{2}}\left( 2{{\gamma }_{rd}}+U \right)}}}  \nonumber\\
 & \!-\!\frac{N}{1\!+\!N}\!\sqrt{\frac{{{P}_{R}}{\gamma }_{rd}}{d_{rd}^{\alpha }}}\rm{erf}\!\left(\!-\sqrt{\!\frac{2N\gamma _{rd}^{2}}{{{(1\!+\!N)}^{2}}\left( 2{{\gamma }_{rd}}\!+\!U \right)}} \right)\!.\
\end{align}
Assuming that ${{Y}_{1}}=\max \left\{ \left| {{Z}_{m}} \right| \right\}~{\rm and}~m=1,2\cdots ,N-1.$ The conditional error probability of Walsh-code detection is measured by
\begin{small}
\begin{align} \label{eq:21func}
{{P}_{\rm{ed}}}(e|{{\gamma }_{rd}})=&\int_{0}^{\infty }{\left[ 1-\Pr \left\{ {{Y}_{1}}\le x \right\} \right]}{{f}_{\text{ }|\text{ }{{Z}_{{\hat{m}}}}\text{ }|\text{ }}}(x)dx\nonumber\\
=&\int_{0}^{\infty }{\left[ 1-\prod\limits_{m=1}^{N-1}{\Pr }\left\{ {{Z}_{m}}\le x \right\} \right]}{{f}_{\text{ }|\text{ }{{Z}_{{\hat{m}}}}\text{ }|\text{ }}}(x)dx\nonumber\\
 =&\frac{1}{\sqrt{2\pi \sigma _{\text{ }\!\!|\!\!\text{ }{{Z}_{{\hat{m}}}}\text{ }\!\!|\!\!\text{ }}^{2}}}\int_{0}^{\infty }\!{\left[1\!-{\!{\left[ \operatorname{erf}\left( \!\frac{x}{\sqrt{2\sigma _{2}^{2}}} \right)\right]}^{N-1}} \right]}\nonumber \\
&\times \left\{ {{e}^{-\frac{{{\left(\!x+\!{{\mu }_{_{\left| {{Z}_{{\hat{m}}}} \right|}}} \right)}^{2}}}{2\sigma _{_{\left|{{Z}_{{\hat{m}}}} \right|}}^{2}}}}\!+\!{{e}\!^{-\frac{{{\left( \!x-\!{{\mu }_{_{\left| {{Z}_{{\hat{m}}}} \right|}}} \right)}^{2}}}{2\sigma _{\left| {{Z}_{{\hat{m}}}}\right|}^{2}}}} \right\}dx.
\end{align}
\end{small}
Let $s=\frac{x}{\sqrt{\sum\limits_{l=1}^{{{L}_{rd}}}{h_{rd,l}^{2}}{{E}_{s}}{{N}_{0,rd}}}}$, then the expression is derived as
\begin{align} \label{eq:22func}
{{P}_{\rm{ed}}}(e|{{\gamma }_{rd}})=&\frac{1}{\sqrt{2\pi \eta }}\int_{0}^{\infty }{\left[ 1-{{\left[ \operatorname{erf}\left( \frac{s}{\sqrt{2\lambda }} \right) \right]}^{N-1}} \right]}\nonumber\\
&\times \left\{ {{e}^{-\frac{{{(v-\Psi )}^{2}}}{2\eta  }}}+{{e}^{-\frac{{{(v+\Psi )}^{2}}}{2\eta  }}} \right\}ds.
\end{align}

Actually, the proposed design criterion is independent of the fading distribution. To illustrate the advantage of the proposed system in a simple and clear way, we consider the multipath Rayleigh fading channel here, as in \cite{7370796,780045,6184294,6560492}. We assume that ${{L}_{\upsilon }}$ channel coefficients are independent and identically distributed random variables over a multipath Rayleigh fading channel. Moreover, we assume that the average power gains of the $L_v$ are equal, i.e., $E\{h_{\upsilon,1}^{2}\}=\ldots =E\{h_{\upsilon, {{L}_{\upsilon }}}^{2}\}$, where $\upsilon \in \{sr,sd,rd\}$. Hence, the PDF of instantaneous SNR ${{\gamma }_{\upsilon }}$ can be written as~\cite{9361588,9758714}
\begin{align} \label{eq:23func}
f({{\gamma }_{\upsilon }})=\frac{\gamma _{_{\upsilon}}^{{{L}_{\upsilon}}-1}}{({{L}_{\upsilon}}-1)!{{{\bar{\gamma }}}^{{{L}_{\upsilon}}}}_{{}}}\exp \left( -\frac{{{\gamma }_{\upsilon}}}{{\bar{\gamma }}} \right),
\end{align}
where $\bar{\gamma } =\frac{{{E}_{s}}}{d_{\upsilon }^{\alpha }{{N}_{0,\upsilon}}}E \{ h_{_{\upsilon,i}}^{2} \}$ is the average SNR, and $i \in \{1,2,\ldots, L_{\upsilon}\}$.

Besides, for ${{L}_{\upsilon }}$ independent Rayleigh fading channels with unequal power delay profiles, the PDF of  ${{\gamma }_{\upsilon }}$ can be written as \cite{8928564}
\begin{align}
f({{\gamma }_{\upsilon }})=\sum\limits_{l=1}^{{{L}_{\upsilon }}}{\frac{\prod\limits_{j=1,j\ne l}^{{{L}_{\upsilon }}}{\left[ \frac{{{{\bar{\gamma }}}_{l}}}{{{{\bar{\gamma }}}_{l}}-{{{\bar{\gamma }}}_{j}}} \right]}}{{{{\bar{\gamma }}}_{l}}}}\exp \left( -\frac{{{\gamma }_{\upsilon }}}{{{{\bar{\gamma }}}_{l}}} \right),
\end{align}
where ${{\bar{\gamma }}_{l}}$ is the average value of ${{\gamma }_{l}}=\frac{h_{\upsilon ,l}^{2}{{E}_{s}}}{d_{\upsilon }^{\alpha }{{N}_{0,\upsilon }}}$, which is the instantaneous SNR on the $l$-th path.

Finally, the average error probability of Walsh-code detection becomes
\begin{align} \label{eq:24func}
{{P}_{\rm{ed}}}=\int_{0}^{\infty }{{{P}_{\rm ed}}(e|{{\gamma }_{rd}})}f({{\gamma }_{rd}})d{{\gamma }_{rd}}.
\end{align}
\subsubsection{Derivation of ${P}_{\rm{{mod}}}$}\label{sect:Derivation of ${P}_{{mod}}$}
The modulated bit transmitted from S and forwarded by R is recovered at D. We first need to analyze the error probability of DF protocol, which is dependent on the erroneous detection of SR-DCSK at R. To get the the decision metric (i.e., ${{Z}_{sr}}$) of SR-DCSK detection, one can get the mean and variance of ${{Z}_{sr}}$, respectively, as
\vspace{-2.5mm}
\begin{subequations}
\begin{align} \label{eq:23func}
E\left\{ {{Z}_{sr}} \right\}&={{b}_{l}}\sum\limits_{l=1}^{{{L}_{sr}}}{h_{sr,l}^{2}}\frac{N{{P}_{S}}{{E}_{s}}}{\left( 1+N \right)d_{sr}^{\alpha }},\\
V\left\{ {{Z}_{sr}} \right\}&=\sum\limits_{l=1}^{{{L}_{sr}}}{h_{sr,l}^{2}}\frac{{N{P}_{S}}{{E}_{s}}{{N}_{0,sr}}}{2d_{sr}^{\alpha }}+NU\frac{N_{0,sr}^{2}}{4}.
\end{align}
\end{subequations}
According to the central limit theorem, the decision metric ${{Z}_{sr}}$ follows the normal distribution~\cite{7370796}, the conditional error probability of DF protocol can be computed as
\vspace{-2.5mm}
\begin{align}\label{eq:24func}
  {{P}_{\text{df}}}(e|{{\gamma }_{sr}})&= \frac{1}{2}\operatorname{erfc}\left( {{\left[ \frac{2V\left\{ {{Z}_{sr}} \right\}}{E{{\left\{ {{Z}_{sr}} \right\}}^{2}}} \right]}^{-\frac{1}{2}}} \right)\text{ }\nonumber \\
 & =\!\frac{1}{2}\operatorname{erfc}\left(\! {{\left[ \frac{{{\left( 1\!+\!N \right)}^{2}}}{N{{\gamma }_{sr}}}\!+\frac{{{(1\!+\!N)}^{2}}U}{2N\gamma _{sr}^{2}} \right]}^{-\frac{1}{2}}}\! \right)\!,
\end{align}
where ${\rm erfc}(x)$ is the complementary error function and ${{\gamma }_{sr}}=\sum\limits_{l=1}^{{{L}_{sr}}}{\frac{h_{sr,l}^{2}{{P}_{S}}{{E}_{s}}}{d_{sr}^{\alpha }{{N}_{0,sr}}}}$ is the instantaneous SNR of S$\rightarrow$R link. The average error probability of DF protocol becomes
\begin{align} \label{eq:28func}
{{P}_{\rm{df}}}=\int_{0}^{\infty }{{{P}_{\rm{df}}}(e|{{\gamma }_{sr}})}f({{\gamma }_{sr}})d{{\gamma }_{sr}}.
\end{align}
At D, we utilize EGC to obtain the decision metric ${{Z}_{egc}}$ of the modulated bit. Actually, the decision metric can be calculated by the those of S$\rightarrow$D and R$\rightarrow$D links, i.e., ${{Z}_{egc}}={\left( {{Z}_{sd}}+{{Z}_{rd}} \right)}/{2}$. When the Walsh-code detection is correct, the decision metric of R$\rightarrow$D link can be written as ${{Z}_{rd}}={{Z}_{{\hat{m}}}}$, the decision metric is expressed as ${{Z}_{egc}}=\left( {{Z}_{sd}}+{{Z}_{{\hat{m}}}} \right)/2$; otherwise, ${{Z}_{rd}}={{Z}_{{m}}}$ ($m\in \{1,2, \cdots ,N\}$ and $m\notin \hat{m}$), the decision metric becomes becomes ${{Z}_{egc}}=\left( {{Z}_{sd}}+{{Z}_{m}} \right)/2$. Thus, the mean and variance of the decision metric conditioned on correct detection at D are respectively given by
\begin{subequations}
\begin{align}
   E\left\{ {{Z}_{egc}} \right\}=&\frac{1}{2}\left( {{b}_{l}}\sum\limits_{l=1}^{{{L}_{sd}}}{h_{sd,l}^{2}}\frac{N{{P}_{S}}{{E}_{s}}}{\left( 1+N \right)d_{sd}^{\alpha }} \right.\nonumber \\
  &\left. +{{b}_{l}}\sum\limits_{l=1}^{{{L}_{rd}}}{h_{rd,l}^{2}}\frac{N{{P}_{R}}{{E}_{s}}}{\left( 1+N \right)d_{rd}^{\alpha }} \right),
\end{align}
\begin{align}
V\left\{ {{Z}_{egc}} \right\}=&\frac{1}{4}\left( \sum\limits_{l=1}^{{{L}_{sd}}}{h_{sd,l}^{2}}\frac{N{{P}_{S}}{{E}_{s}}{{N}_{0,sd}}}{2d_{sd}^{\alpha }} \right.\nonumber \\
 & +\sum\limits_{l=1}^{{{L}_{rd}}}{h_{rd,l}^{2}}\frac{N{{P}_{R}}{{E}_{s}}{{N}_{0,rd}}}{2d_{rd}^{\alpha }}\nonumber \\
 & +NU\frac{N_{0,sd}^{2}}{4}+\left. NU\frac{N_{0,rd}^{2}}{4} \right).
\end{align}
\end{subequations}

On the contrary, the mean and variance of the decision metric conditioned on erroneous detection at D are respectively give by
\begin{subequations}
\begin{align} \label{eq:31func}
E\left\{ {{Z}_{egc}} \right\}=&\frac{1}{2}\left( {{b}_{l}}\sum\limits_{l=1}^{{{L}_{sd}}}{h_{sd,l}^{2}}\frac{N{{P}_{S}}{{E}_{s}}}{\left( 1+N \right)d_{sd}^{\alpha }} \right),\nonumber\\
   V\left\{ {{Z}_{egc}} \right\}=&\frac{1}{4}\left( \sum\limits_{l=1}^{{{L}_{sd}}}{h_{sd,l}^{2}}\frac{N{{P}_{S}}{{E}_{s}}{{N}_{0,sd}}}{2d_{sd}^{\alpha }} \right. \\
 & +\sum\limits_{l=1}^{{{L}_{rd}}}{h_{rd,l}^{2}}\frac{N{{P}_{R}}{{E}_{s}}{{N}_{0,rd}}}{2(1+N)d_{rd}^{\alpha }} \nonumber\\
 & +NU\frac{N_{0,sd}^{2}}{4}\left. +NU\frac{N_{0,rd}^{2}}{4} \right).
\end{align}
\end{subequations}

For the BER calculation of modulated bit, there are four possible cases occurring in CIM-SR-DCSK-CC system. Specifically, the four cases are defined as follows:

1) {\bf Case 1~$\bm{(\Theta =1)}$:} 
Both the SR-DCSK detection at R and the Walsh-code detection at D are correct;

2) {\bf Case 2~$\bm{(\Theta =2)}$:}  The SR-DCSK detection at R is incorrect, but the Walsh-code detection at D is correct;

3) {\bf Case 3~$\bm{(\Theta =3)}$:}  The SR-DCSK detection at R is correct, but the Walsh-code detection at D is incorrect;

4) {\bf Case 4~$\bm{(\Theta =4)}$:} Both the SR-DCSK detection at R and the Walsh-code detection at D are incorrect.

As such, the BER of modulated bit can be calculated as
\begin{align} \label{eq:33func}
{{P}_{\rm{mod}}}=\sum\limits_{\Lambda =1}^{4}{\Pr \left( \Theta =\Lambda  \right)P(e|\Theta =\Lambda )},
\end{align}
where $\Pr \left( \Theta =1 \right)$, $\Pr \left( \Theta =2 \right)$, $\Pr \left( \Theta =3 \right)$, and $\Pr \left( \Theta =4 \right)$ represent the occurrence probabilities of case 1, case 2, case 3, and case 4, respectively.
$P({e| \Theta =1 })$, $P({e| \Theta =2 })$, $P({e| \Theta =3 })$, and $P({e| \Theta =4 })$ represent the conditional BERs in case 1, case 2, case 3, and case 4, respectively. In particular, the conditional BERs of the latter two are equal, i.e., $P({e| \Theta =3 })=P({e| \Theta =4 })$, because their decision metrics obey the same normal distribution. On the one hand,
the occurrence probabilities of four cases are computed as
\begin{subequations}\label{eq:34func}
 \begin{align}
 &\Pr \left( \Theta =1 \right)=(1-{{P}_{\rm df}})(1-{{P}_{\rm ed}}),\\
 &\Pr \left( \Theta =2 \right)={{P}_{\rm df}}(1-{{P}_{\rm ed}}),\\
  &\Pr \left( \Theta =3 \right)={{P}_{\rm ed}}(1-{{P}_{\rm df}}),\\
  &\Pr \left( \Theta =4 \right)={{P}_{\rm ed}}{{P}_{\rm df}}.
 \end{align}
\end{subequations}
On the other hand, the conditional BERs of four cases are calculated in (\ref{eq:35func}), as shown at the top of the next page, where ${{\gamma }_{sd}}=\sum\limits_{l=1}^{{{L}_{sd}}}{\frac{h_{sd,l}^{2}{{P}_{S}}{{E}_{s}}}{d_{sd}^{\alpha }{{N}_{0,sd}}}}$ is the instantaneous SNR of S$\rightarrow$D link.
As such, the BER of modulated bit can be finally derived by substituting Eq.~(\ref{eq:34func}) and Eq.~(\ref{eq:35func}) into  Eq.~(\ref{eq:33func}).
\begin{figure*}[htbp]
\begin{subequations} \label{eq:35func}
\begin{align} \label{eq:34afunc}
P({e| \Theta =1 })&=\frac{1}{2}\int_{0}^{\infty }{\int_{0}^{\infty }{}\operatorname{erfc}\left( {{\left[ \frac{\left( 1+N \right)}{\sqrt{N}}\frac{\sqrt{{{\gamma }_{sd}}+{{\gamma }_{rd}}+U}}{\left( {{\gamma }_{sd}}+{{\gamma }_{rd}} \right)} \right]}^{-1}} \right)f\left( {{\gamma }_{sd}} \right)f\left( {{\gamma }_{rd}} \right)d{{\gamma }_{sd}}d{{\gamma }_{rd}}},\\
P({e| \Theta =2 })&=\frac{1}{2}\int_{0}^{\infty }{\int_{0}^{\infty }{}\operatorname{erfc}\left( {{\left[ \frac{\left( 1+N \right)}{\sqrt{N}}\frac{\sqrt{{{\gamma }_{sd}}+{{\gamma }_{rd}}+U}}{\left( {{\gamma }_{sd}}-{{\gamma }_{rd}} \right)} \right]}^{-1}} \right)f\left( {{\gamma }_{sd}} \right)f\left( {{\gamma }_{rd}} \right)d{{\gamma }_{sd}}d{{\gamma }_{rd}}},\\
P({e| \Theta =3 })&=P({e| \Theta =4 })=\frac{1}{2}\!\int_{0}^{\infty }{\!\int_{0}^{\infty }{\operatorname{erfc}}\left( {{\left[ \frac{\!\left( 1+N \right)}{\sqrt{N}}\frac{\sqrt{{{\gamma }_{sd}}+\!\frac{{{\gamma }_{rd}}}{\!\left( N+1\! \right)}+\!U}}{{{\gamma }_{sd}}} \right]}^{-1}} \right)\!f\left( {{\gamma }_{sd}} \right)\!f\left( {{\gamma }_{rd}} \right)d{{\gamma }_{sd}}d{{\gamma }_{rd}}},
\end{align}
\vspace{-9mm}
\end{subequations}
\end{figure*}

Based on the resultant ${{P}_{\rm cim}}$ and ${{P}_{\rm mod}}$, one can promptly get the system BER ${{P}_{\rm sys}}$ over a multipath Rayleigh fading channel by exploiting Eq.~\eqref{eq:12func}.
Furthermore, when the number of paths ${{L}_{\upsilon }}=1$ and the channel coefficient ${{h}_{\upsilon,1 }}=1$, the multipath Rayleigh fading channel reduces to an AWGN channel. In this sense, the system BER over an AWGN channel can be readily obtained.

\subsection{Transmission Throughput}\label{Throughput analysis}
As in \cite{8928564}, we define that the normalized throughput is the ratio of the successfully received bits/second of a given DCSK system to the transmitted bits/second of the proposed CIM-SR-DCSK-CC system.
According to Sect.~\ref{sect: a CIM-SR-DCSK-CC system model}, the proposed CIM-SR-DCSK-CC system can transmit $N_p$ information bit streams in each transmission period (i.e, $\bar{T}={\left( U+\beta \right){{N}_{p}}{{T}_{c}}\times 2}$).

Therefore, the normalized throughput can be formulated as
\begin{align}\label{eq:39func}
{{R}_{t}}= \frac{[{{\left( 1-P_{\rm sys}^{t} \right)}^{{{N}_{p}}}} N_p]/{{{{\bar{T}}}_{t}}}} {N_p / \bar{T}}
=\frac{{{\left( 1-P_{\rm sys}^{t} \right)}^{{{N}_{p}}}}\bar{T}}{{{{\bar{T}}}_{t}}},
\end{align}
where $P_{\rm sys}^{t}$ and ${{\bar{T}}_{t}}$ denote the average BER  and time duration required to transmit $N_p$ information bit streams for a given system $t\in \left\{ \text{DCSK-CC},\text{SR-DCSK-CC},\text{CIM-SR-DCSK-CC} \right\}$, respectively.\footnote{For comparison, we establish an SR-DCSK-CC system by combining the SR-DCSK \cite{7370796} with  cooperative communication technique as a new baseline.}\  Actually, the DCSK-CC system \cite{8036271,9434904} and SR-DCSK-CC system need three time slots to transmit $N_p$ information bit streams to the destination, but the proposed CIM-SR-DCSK-CC system only needs two time slots to achieve the same goal. Hence, the time durations required to transmit $N_p$ information bit streams for DCSK-CC, SR-DCSK-CC and CIM-SR-DCSK-CC are ${{\bar{T}}_{\text {DCSK-CC}}}={2\beta {{N}_{p}}{{T}_{c}}\times 3}$, ${{\bar{T}}_{\text {SR-DCSK-CC}}}={\left( U+ \beta\right){{N}_{p}}{{T}_{c}}\times 3}$ and ${{\bar{T}}_{\text{ CIM-SR-DCSK-CC}}}={\left( U+\beta \right){{N}_{p}}{{T}_{c}}\times 2}$, respectively.

Based on the above discussion, we will compare the normalized throughput among the proposed CIM-SR-DCSK-CC system, DCSK-CC system, and SR-DCSK-CC system in the forthcoming section to illustrate the advantage of our design.

\section{Simulation Results and Discussions}\label{sect:simulation results and discussion}
\vspace{-2mm}
In this section, a variety of simulations are carried out to evaluate the performance of the proposed CIM-SR-DCSK-CC system over AWGN and multipath Rayleigh fading channels, which verify the theoretical derivations in the previous section. We also compare the performance of the proposed system with the existing DCSK-CC system \cite{8036271,9434904} and SR-DCSK-CC system to illustrate the advantage of our design. Unless otherwise mentioned, in the simulations, we set the spreading factor as $SF=510$, the transmit power of S and R as ${{P}_{S}}={{P}_{R}}=1\rm W$. Without loss of generality, we assume that the path loss coefficient
$\alpha=2$ \cite{5629387,7577838,6197159}, the noise variances of S$\rightarrow$R link, R$\rightarrow$D link, and S$\rightarrow$D link as ${{N}_{0,sr}}={{N}_{0,rd}}={{N}_{0,sd}}={{N}_{0}}$, and their corresponding distances as ${{d}_{sr}}={{d}_{rd}}=1{\rm m},{{d}_{sd}}=2{\rm m}$.
For the multipath Rayleigh fading channel, we consider three component paths with the following parameter setting: the average power gains are $E\{h_{\upsilon,1}^{2}\}=E\{h_{\upsilon,2}^{2}\} =E\{h_{\upsilon,3}^{2}\}=1/3$, and the path delays are ${{\tau }_{\upsilon,1}}=0,{{\tau }_{\upsilon,2}}={{T}_{c}},{{\tau }_{\upsilon,3}}=2{{T}_{c}}$, where $\upsilon \in \{sr,sd,rd\}$.
\begin{figure*}[htbp]
\centering
\subfigure[]{
\includegraphics[width=3.25in,height=2.5in]{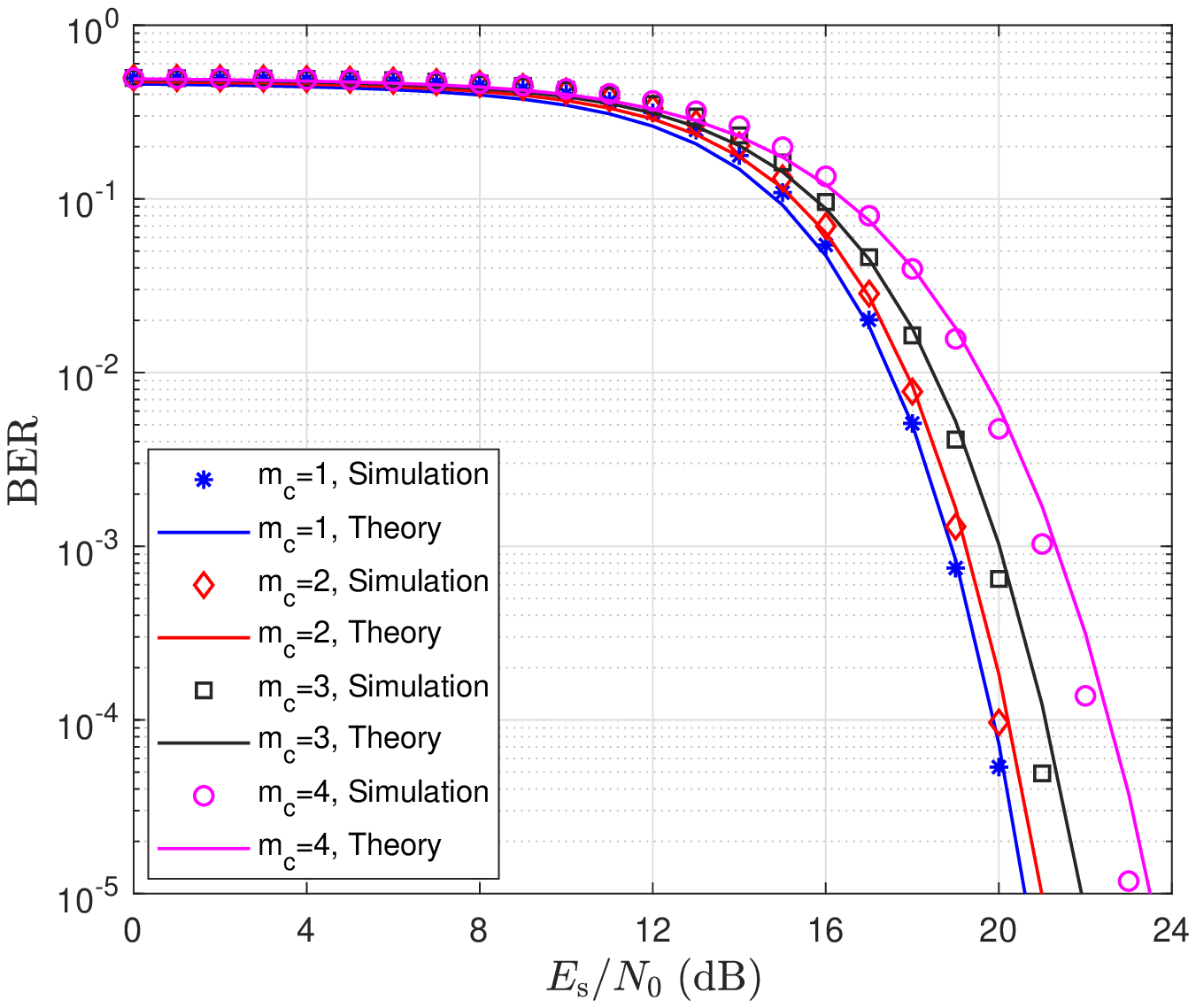}} \hspace{6mm}
\subfigure[]{
\includegraphics[width=3.25in,height=2.5in]{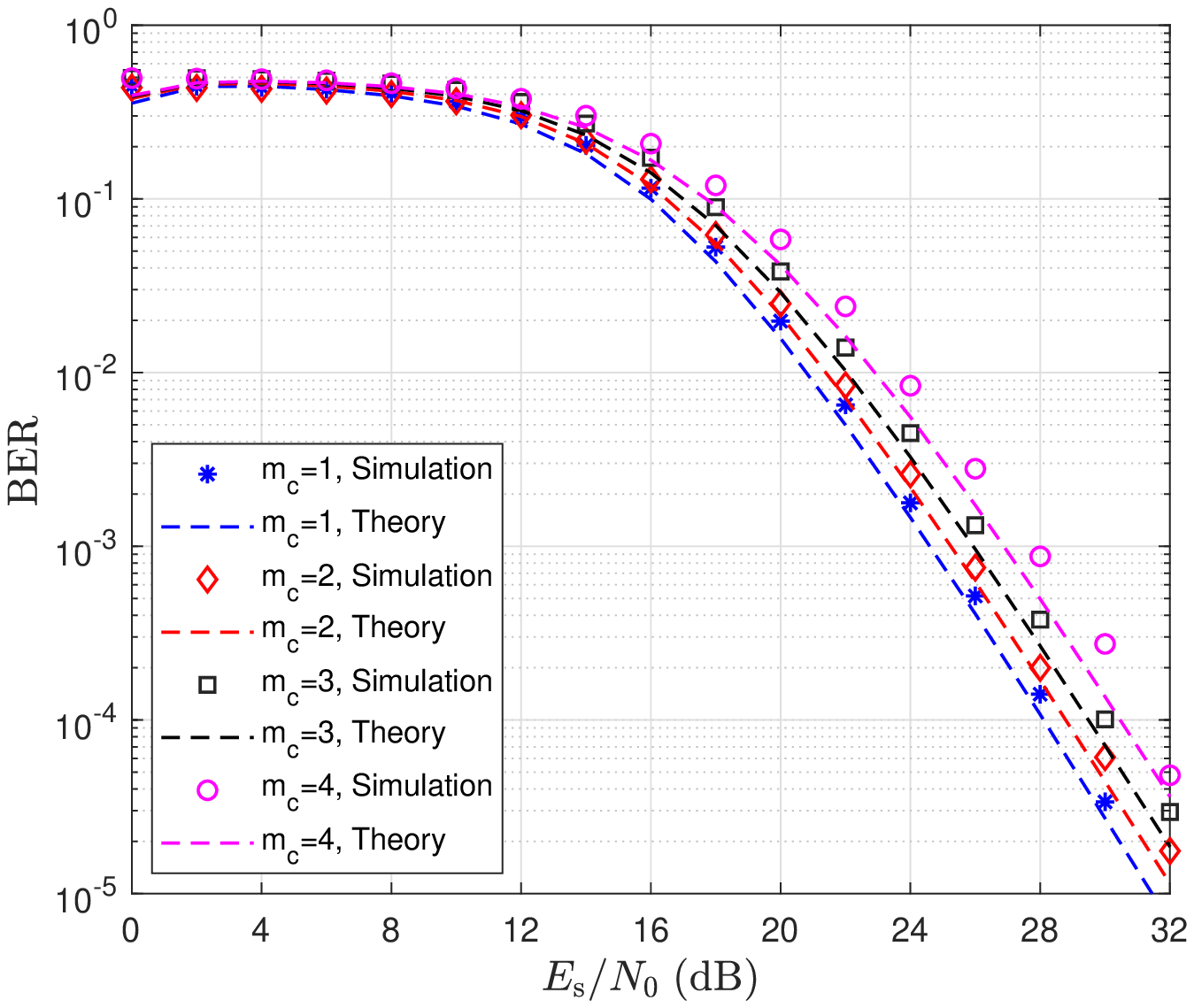}}
\caption{Simulated and theoretical BER performance of the proposed CIM-SR-DCSK-CC system over (a) AWGN and (b) multipath Rayleigh fading channels, where ${m}_{c}=1,2,3,4$.}
\label{fig:fig5}
\end{figure*}
\subsection{Verification of Theoretical BER Analysis}\label{sect:Verification}

Figure~\ref {fig:fig5} shows the simulated and theoretical BER performance of the proposed CIM-SR-DCSK-CC system over AWGN and multipath Rayleigh fading channels.
Obviously, we can see that the simulated and theoretical BER curves are well consistent with each other. Besides, the BER performance of the proposed system gets worse with the increase of ${m}_{c}$.
Actually, the system can allow the relay to transmit additional information bits at the price of sacrificing some BER performance. As a result, one should strike a balance between the error performance and data rate for the CIM-SR-DCSK-CC systems in practical applications.

\subsection{Discussion on the Influence of Placement of R on BER performance}\label{sect:Verification}

\begin{figure}[h]
\centering
\includegraphics[width=3.25in,height=2.5in]{{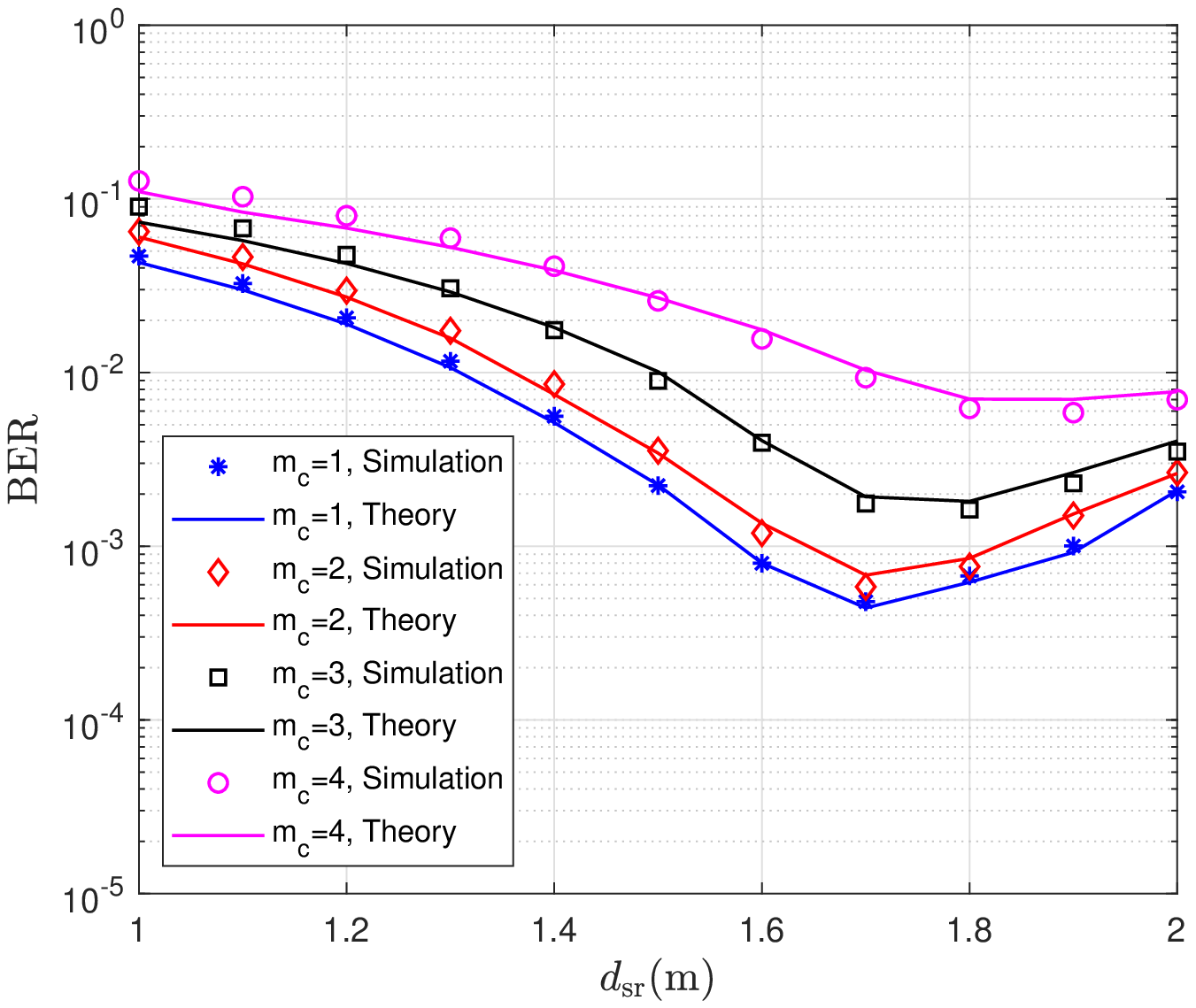}}
\caption{Simulated and theoretical BER performance of the proposed CIM-SR-DCSK-CC system versus ${{d}_{sr}}$ over an AWGN channel, where~${{E}_{s}}/{{N}_{0}}=22~\rm{dB}$ and ${m}_{c}=1,2,3,4$.}
\label{fig:fig6}  
\end{figure}

As a further advance, we investigate the influence of the distance ${{d}_{sr}}$ between S and R (resp. ${{d}_{rd}}$) on the BER performance of the proposed CIM-SR-DCSK-CC system over an AWGN channel when ${E}_{s}/{N}_{0}=22~{\rm dB}$, as shown in Figure~\ref {fig:fig6}. In this figure, we assume that the distance between S and D is ${{d}_{sd}}=3\rm m$, while ${{d}_{sr}}$ is varying from $1{\rm m}$ to $2{\rm m}$ (i.e., ${{d}_{rd}}$ is varying from $2{\rm m}$ to $1{\rm m}$). As seen, for a given number of index bits (i.e., a given value of ${m}_{c}$), the BER performance of the CIM-SR-DCSK-CC system first improves and then deteriorates as ${{d}_{sr}}$ increases.
It is because that as R moves away from S (i.e., R gets closer to D), the BER of system is mainly dependent on the BER of index bits in the initial stage, but not the BER of the modulated bit (see Eq.~\eqref{eq:12func}).
Nevertheless, once ${{d}_{sr}}$ exceeds a certain value, the BER of modulated bit, instead of the BER of index bits, plays an more important role on the BER of system.
In this stage, the probability of successful decoding at R is greatly reduced, leading to a severe degradation in the BER of modulated bit. Hence, the BER of system becomes worse in this stage. Besides, one can observe that the simulated BER curves agree well with the theoretical ones as ${{d}_{sr}}$ varies, which further validate the accuracy of theoretical BER derivation. Simulations have also performed over a multipath Rayleigh fading channel and similar observations have been obtained.

\subsection{BER Comparison Among the Proposed CIM-SR-DCSK-CC System, DCSK-CC System, and SR-DCSK-CC System}\label{sect:Verification}

In the proposed CIM-SR-DCSK-CC system, both the source and relay can transmit their own information bits to the destination separately. However, in the two baselines (i.e., the DCSK-CC system and SR-DCSK-CC system), only the source transmits its own information bits to the destination. Hence, to guarantee a fair performance comparison, we consider the following transmission mechanism for the above two DCSK-CC systems. In the considered mechanism, the DCSK-CC system and SR-DCSK-CC system require two time slots to transmit the source information bits and spend an additional time slot to transmit the relay information bits.
Since the relay in either the DCSK-CC system or SR-DCSK-CC system can transmit only one information bit in the third time slot, we also assume that in the proposed CIM-SR-DCSK-CC system the relay transmits one index bit (i.e., ${m}_{c}=1$) for the sake of fairness. Moreover, we assume that the total transmission energy $E_{\rm T}$ for all the three systems are identical.

\begin{figure}[htbp]
\centering
\subfigure[]{ \label{fig:subfig7a}
\includegraphics[width=3.25in,height=2.5in]{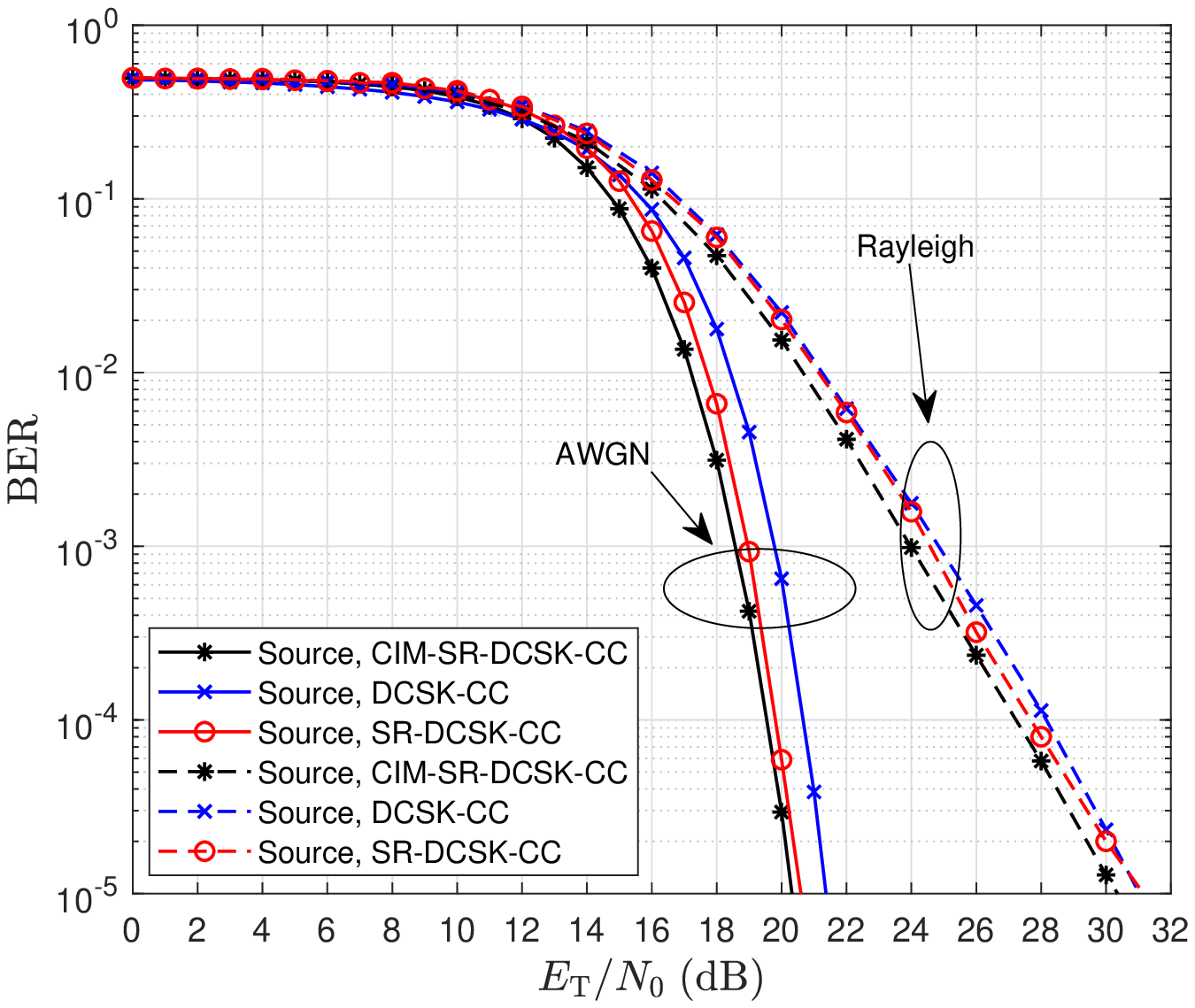}}\hspace{-6.5mm}
\subfigure[]{ \label{fig:subfig7b}
\includegraphics[width=3.25in,height=2.5in]{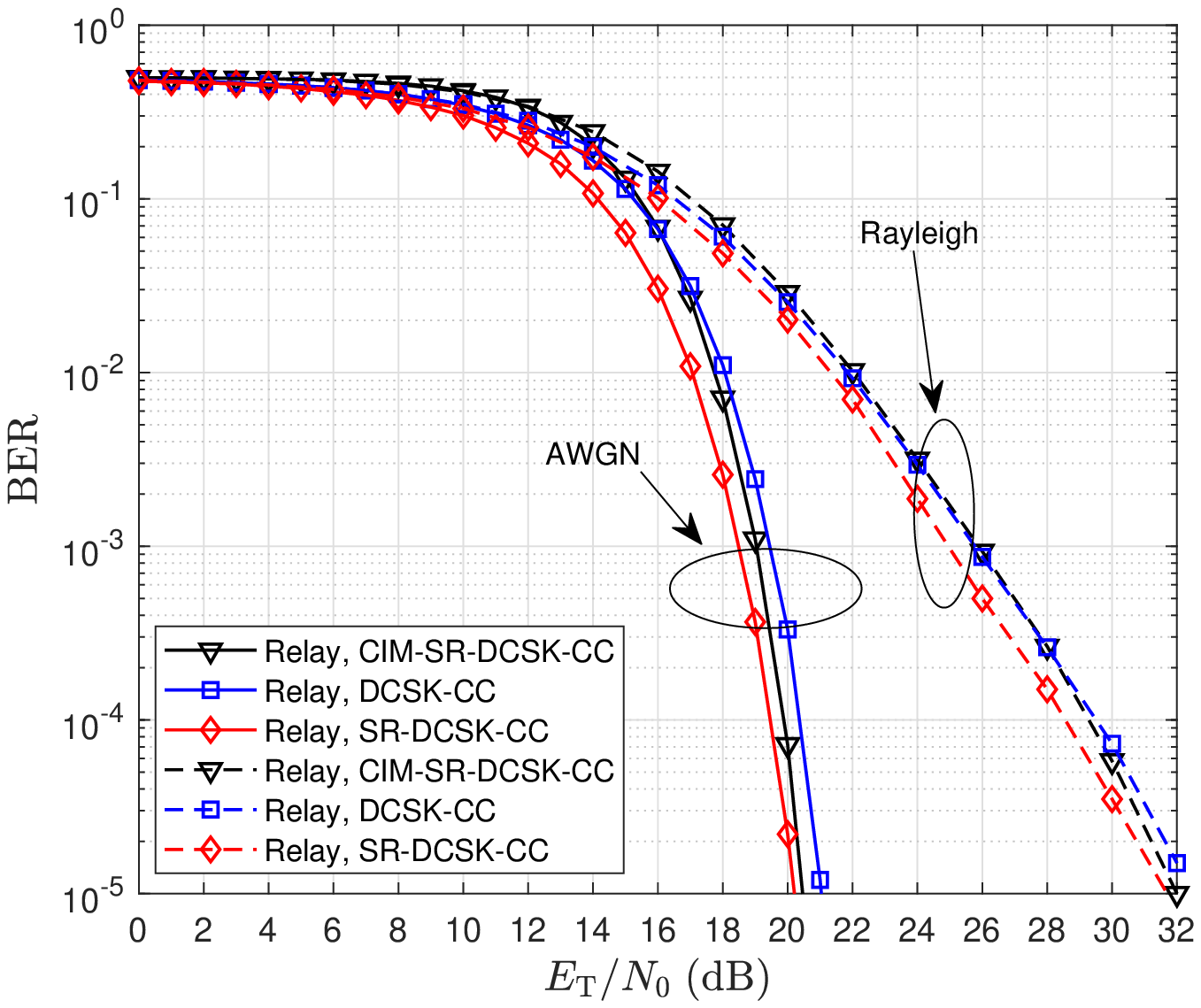}}\vspace{-2mm}
\caption{ BER performance of the (a) source information bits and (b) relay information bits in the proposed CIM-SR-DCSK-CC system, the DCSK-CC system, and the SR-DCSK-CC system over AWGN and multipath Rayleigh fading channels.}
\label{fig:fig7}\vspace{-2mm}
\end{figure}

Based on the above assumption, Figure~\ref{fig:fig7} shows the BER performance of the source information bits and relay information bits in the proposed CIM-SR-DCSK-CC system, the DCSK-CC system and the SR-DCSK-CC system over AWGN and multipath Rayleigh fading channels. In Figure~\ref{fig:fig7}\subref{fig:subfig7a}, it can be seen that the BER performance of the source information bit in the proposed CIM-SR-DCSK-CC system is better than those in DCSK-CC system and SR-DCSK-CC system over AWGN and multipath Rayleigh fading channels. This is because that the relays in the DCSK-CC system and SR-DCSK-CC system needs some energy to transmit their own information bits, while
the relay in the CIM-SR-DCSK-CC system does not need additional energy to transmit its own information bit. As a result, the CIM-SR-DCSK-CC system has more energy to transmit the source information bits with respect to the other two counterparts.
Referring to Figure~\ref{fig:fig7}\subref{fig:subfig7b}, the CIM-SR-DCSK-CC system has better BER performance of the relay information bit than the DCSK-CC system, but worse than SR-DCSK-CC system over both channels. This phenomenon is due to the fact that the BER of relay information bit in the CIM-SR-DCSK-CC system is affected by the Walsh-code detection compared to SR-DCSK-CC system.
\begin{figure}[tbp]
\center
\includegraphics[width=3.4in,height=2.6in]{{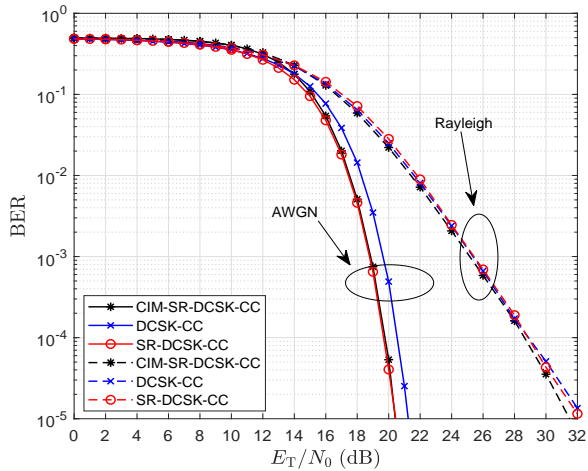}}
\caption{Overall BER performance of the proposed CIM-SR-DCSK-CC system, the DCSK-CC system, and the SR-DCSK-CC system over AWGN and multipath Rayleigh fading channels.}
\label{fig:fig8}  
\vspace{-2mm}
\end{figure}

Figure~\ref{fig:fig8} shows the overall BER performance of the proposed CIM-SR-DCSK-CC system, DCSK-CC system and SR-DCSK-CC system over AWGN and multipath Rayleigh fading channels. The BER performance of the proposed CIM-SR-DCSK-CC system is similar to that of the SR-DCSK-CC system, both of which are better than that of DCSK-CC system over an AWGN channel. In addition, the BER performance of the CIM-SR-DCSK-CC system is slightly better than that of DCSK-CC system and SR-DCSK-CC system over a multipath Rayleigh fading channel. For example, the proposed CIM-SR-DCSK-CC system achieves a gain of $0.5$~dB compared with the DCSK-CC system at a BER of \begin{math}10^{-5}\end{math} over AWGN and multipath Rayleigh fading channels. Actually, the BER performance of CIM-SR-DCSK-CC system is better than that of DCSK-CC system from both the source-information-bit and relay-information-bit perspectives.

\subsection{Throughput Comparison Among the Proposed CIM-SR-DCSK-CC System, DCSK-CC System, and SR-DCSK-CC System}\label{sect:Verification}
\begin{figure}[htpb]
\centering
\subfigure[\hspace{-0.5cm}]{
\includegraphics[width=3.4in,height=2.6in]{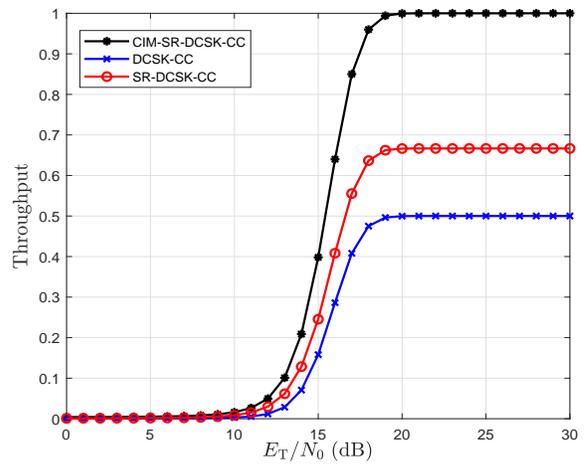}}\hspace{-6.5mm}
\subfigure[\hspace{-0.5cm}\vspace{-3mm}]{
\includegraphics[width=3.4in,height=2.6in]{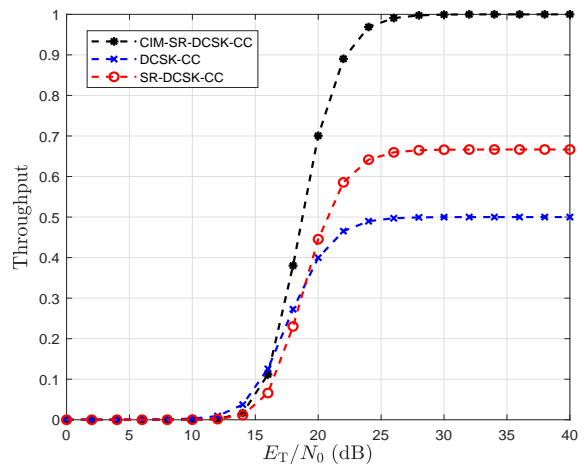}}
\caption{ Normalized throughputs of the proposed CIM-SR-DCSK-CC system, the SR-DCSK-CC system, and the DCSK-CC system over (a) AWGN and (b) multipath Rayleigh fading channels.}
\label{fig:fig9}\vspace{-3mm}
\end{figure}
The normalized throughputs of the proposed CIM-SR-DCSK-CC system, the DCSK-CC system, and the SR-DCSK-CC system over AWGN and multipath Rayleigh fading channels are presented in Figure~\ref{fig:fig9}. It can be observed that the normalized throughput of the proposed CIM-SR-DCSK-CC system is much higher than those of the DCSK-CC system and SR-DCSK-CC system. For example, at $E_{\rm T}/N_0=30~{\rm dB}$, the normalized throughput of the proposed CIM-SR-DCSK-CC system is $1$ over AWGN and multipath Rayleigh fading channels, while those of the DCSK-CC system and SR-DCSK-CC system are $1/2$ and $2/3$, respectively. This is because that the proposed system has a great improvement in time duration for information transmission with respect to the other two counterparts.
{\em Remark:}
Although the aforementioned BER results are obtained under a given parameter setting, simulations have also been carried out with other parameter settings (i.e., different $SF, P_S, P_R$, $d_v, \alpha, E\{h_{\upsilon,l}^{2}\},~{\rm and}~\tau_{v,l}$) to substantially verify the advantage of the proposed CIM-SR-DCSCK-CC system.

\section{Conclusion}\label{sect:Conclusions}
In this paper, we put forward a code-index-modulation-aided SR-DCSK-CC system, namely CIM-SR-DCSK-CC system, to achieve high-throughput transmissions. In the proposed system, the relay not only forwards the source information bit to the destination, but also employs a Walsh code to convey additional information bits simultaneously.
As a further advance, we  analyzed the theoretical BER expressions and throughput of the proposed CIM-SR-DCSK-CC system over AWGN and multipath Rayleigh fading channels, which are in good agreement with the simulation results. Compared with the DCSK-CC system and SR-DCSK-CC system, the proposed CIM-SR-DCSK-CC system exhibits significant improvements in throughput without sacrificing any BER performance. Owing to the appealing advantages, the proposed CIM-SR-DCSK-CC system can be expected to be a competitive alternative for the 6G-enabled low-power and high-rate applications.



\bibliographystyle{IEEEtran}  
\bibliography{IEEEabrv,reff}

\end{document}